\documentclass[prl,10pt,twocolumn,superscriptaddress,notitlepage,floatfix,amssymb]{revtex4}
\usepackage[letterpaper,hmargin={1.5cm,1.5cm},vmargin={1.5cm,2.5cm}]{geometry}
\usepackage{mathtools}
\usepackage{float}
\usepackage{amsfonts}
\usepackage{enumitem}
\usepackage{graphicx}
\usepackage{color}
\newcommand*{\ham}{\hat{H}}

\newcommand*{\cre}[2][a]{\hat{#1}_{#2}^{\dagger}}	

\newcommand*{\ann}[2][a]{\hat{#1}_{#2}}

\newcommand*{\Inp}[1]{\hat{I}_{#1}}
\newcommand*{\Qua}[1]{\hat{Q}_{#1}}

\begin{document}
\title{Generating Multimode Entangled Microwaves with a Superconducting Parametric Cavity}

\author{C.W. Sandbo Chang}
\affiliation{Institute for Quantum Computing and Electrical and Computer Engineering, University of Waterloo, Waterloo, Canada}
\author{M. Simoen}
\affiliation{MC2, Chalmers University of Technology, G\"oteborg, Sweden}
\author{Jos{\'e} Aumentado}
\affiliation{National Institute of Standards and Technology, 325 Broadway, Boulder, Colorado 80305, USA}
\author{Carlos Sab{\'\i}n}
\affiliation{Instituto de F{\'i}sica Fundamental, CSIC, Serrano, 113-bis, 28006 Madrid, Spain}
\author{P. Forn-D\'{i}az}
\affiliation{Institute for Quantum Computing and Electrical and Computer Engineering, University of Waterloo, Waterloo, Canada}
\author{A.M. Vadiraj}
\affiliation{Institute for Quantum Computing and Electrical and Computer Engineering, University of Waterloo, Waterloo, Canada}
\author{Fernando Quijandr\'{\i}a}
\affiliation{MC2, Chalmers University of Technology, G\"oteborg, Sweden} 
\author{G. Johansson}
\affiliation{MC2, Chalmers University of Technology, G\"oteborg, Sweden}
\author{I. Fuentes}
\affiliation{School of Mathematical Sciences, University of Nottingham, Nottingham NG7 2RD, United Kingdom}
\affiliation{Faculty of Physics, University of Vienna, Boltzmanngasse 5, 1090 Vienna, Austria}
\author{C.M. Wilson}
\affiliation{Institute for Quantum Computing and Electrical and Computer Engineering, University of Waterloo, Waterloo, Canada}

\email{chris.wilson@uwaterloo.ca}

%
%

\begin{abstract}
In this Letter, we demonstrate the generation of multimode entangled states of propagating microwaves.  The entangled states are generated by parametrically pumping a multimode superconducting cavity. By combining different pump frequencies, applied simultaneously to the device, we can produce different entanglement structures in a programable fashion.  The Gaussian output states are fully characterized by measuring the full covariance matrices of the modes. The covariance matrices are absolutely calibrated using an \textit{in situ} microwave calibration source, a shot noise tunnel junction. Applying a variety of entanglement measures, we demonstrate both full inseparability and genuine tripartite entanglement of the states.  Our method is easily extensible to more modes. 
\end{abstract}

\date{\today}


\maketitle

The generation and distribution of entanglement is an important problem in quantum information science. For instance, distributing entangled photons is a key paradigm in quantum communication \cite{Jennewein:2000ke}.  Distributing entangled photons as a way to entangle remote processing nodes of a larger quantum computer is also a promising path towards scalability \cite{Kimble:2008if,Felicetti:2014dz}. Multimode entangled states can also be used for a variety of quantum networking protocols such as quantum state sharing \cite{Lance:2004do}, quantum secret sharing \cite{Cleve:1999cb,Tyc:2002cf}, and quantum teleportation networks \cite{Yonezawa:2004wm}. It is therefore of great interest to develop novel ways of efficiently generating propagating entangled states. In this letter, we present a microwave circuit, a multimode parametric cavity, that generates propagating bipartite and tripartite entangled states of microwave photons. Furthermore, the entanglement structure of the tripartite states can be changed \textit{in situ} by the appropriate choice of pump frequencies.  The design is easily extensible to more modes using the same principle and techniques.  The ability to generate complex multimode states with a programmable entanglement structure would potentially enable a number of interesting advances beyond those already mentioned such as microwave cluster states \cite{Bruschi:2016hu}, error-correctable logical qubits for quantum communication \cite{Bolt:2016cv,Jouguet:2013df}, and the quantum simulation of relativistic quantum information processing systems \cite{Bruschi:2013cq, Wilson:2011ir}. 

Superconducting parametric cavities have shown great promise as a quantum technology platform in recent years. Quantum-limited parametric amplifiers have become almost commonplace in superconducting quantum computation. The parametric generation of bipartite continuous-variable (CV) entanglement between two microwave modes has been demonstrated using parametric cavities \cite{Flurin:2012hq,Flurin:2015jf,Fedorov:2016kp,Fedorov:2017tk}. Other work has shown that parametric processes can coherently couple microwave signals between different modes of a single cavity or multiple cavities, including generating superposition states of a single photon at different frequencies \cite{Sirois:2015gr,ZakkaBajjani:2011in}.  The generation of multimode CV entangled states at optical frequencies has also been demonstrated in a variety of ways \cite{Aoki:2003is,Yonezawa:2004wm,Lance:2004do,Pysher:2011hn,Gerke:2016hu,Shalm:2012iaa}.

In this work, by using a multimode superconducting parametric cavity, we demonstrate the generation, calibration and verification of multimode CV entanglement, observing genuine tripartite entanglement of three propagating microwave modes. The states are fully characterized by measuring the 6-by-6 covariance matrix of the mode quadratures. The device operates in steady-state, functioning as a continuous-wave source of entanglement. Our scheme can be extended beyond three modes by simply adding more pump tones.
 

\begin{figure}
\includegraphics[width=0.6\linewidth]{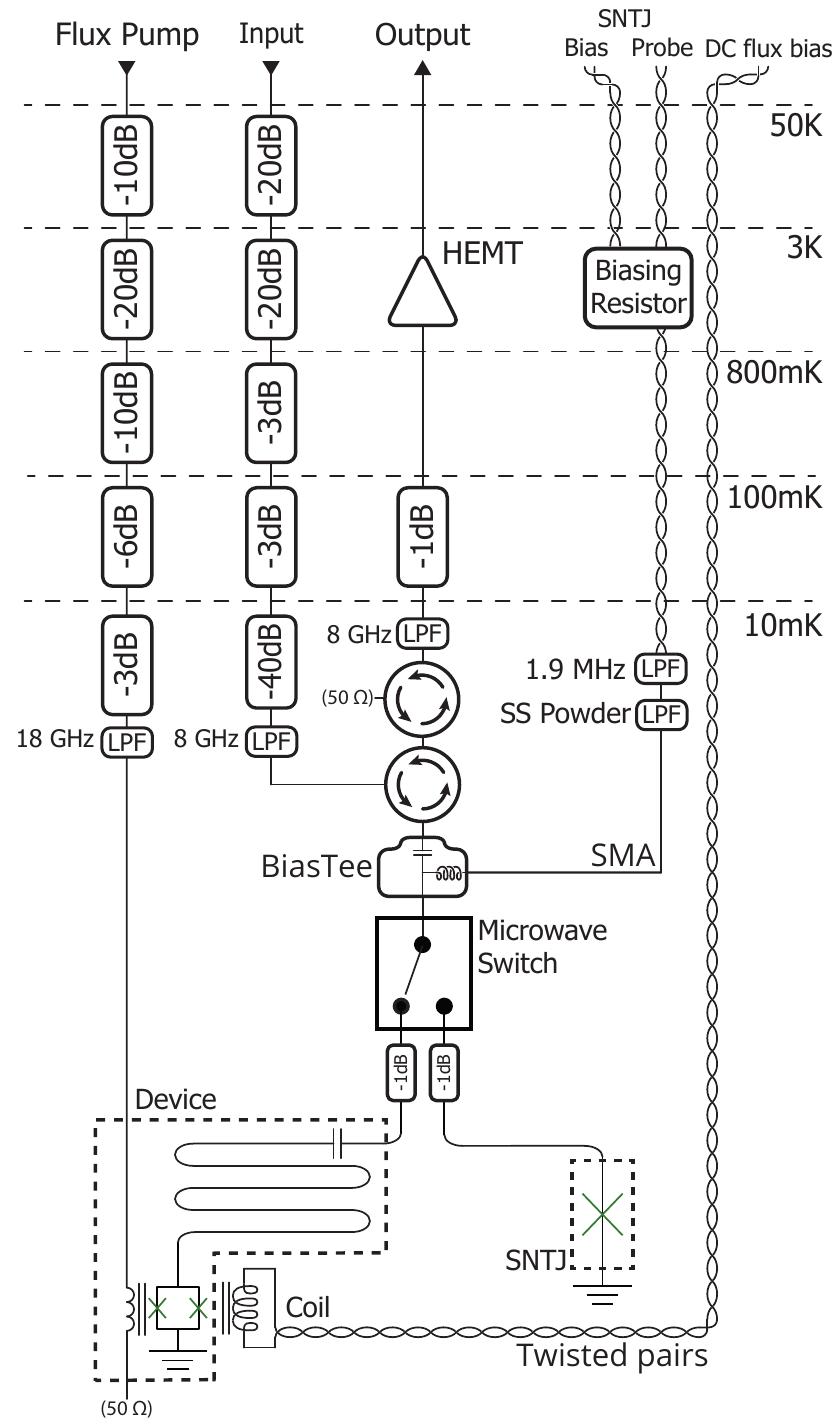}	
\includegraphics[width=0.9\linewidth]{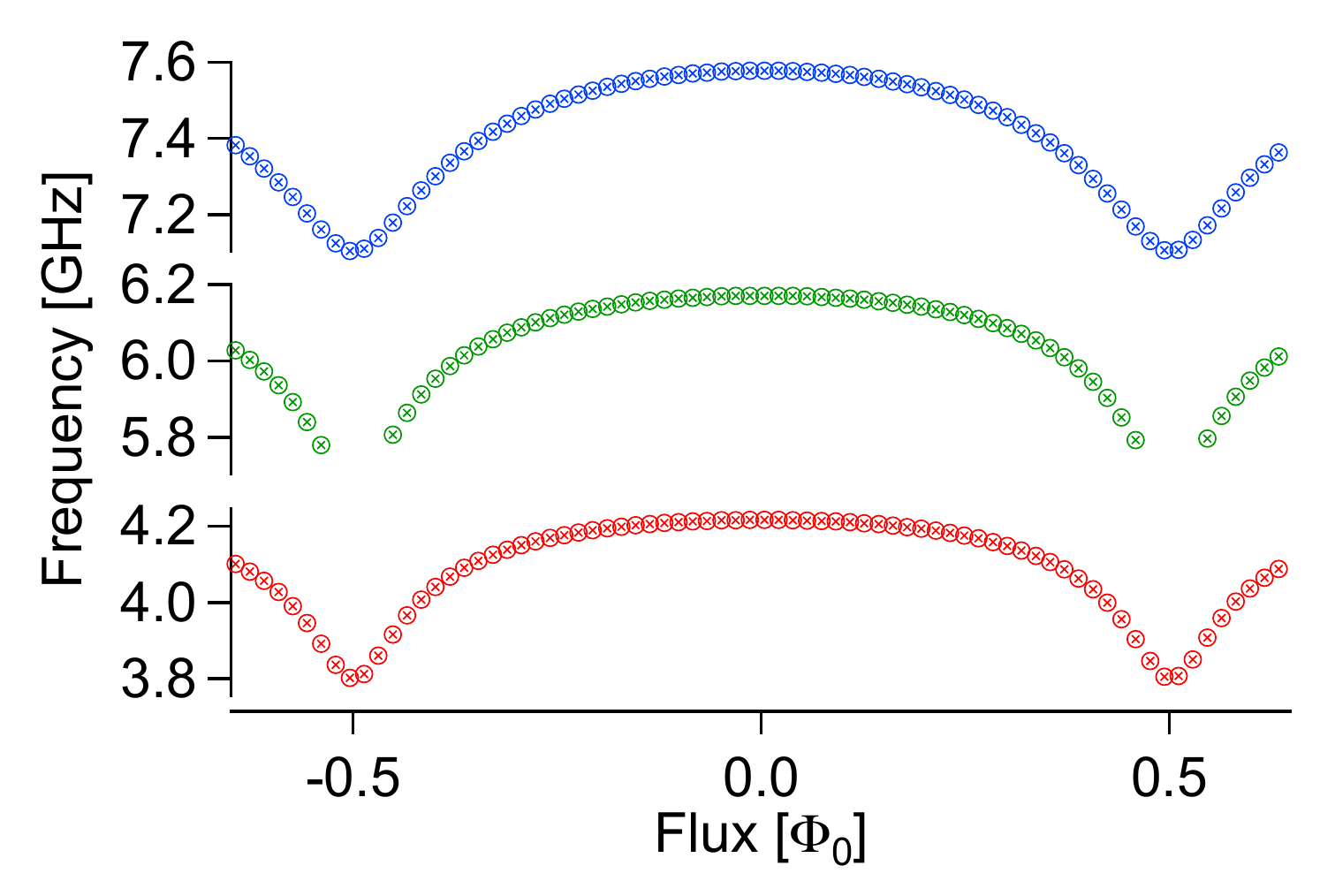}	
\label{Circuit}
\caption{Top Panel: Simplified schematic of the microwave measurement setup. From the bias tee onward, the measurement chain is shared by the parametric cavity and the SNTJ calibration source. The short connections between the switch and two devices are made as physically identical as possible. The system is calibrated independently at each of the measurement frequencies. Bottom Panels: Tuning curves of the three modes of the cavity, showing the tuning of the measured resonance frequencies with external magnetic flux, $\Phi_{\text{ext}}$ (in units of the flux quantum $\Phi_0$). The maximum frequency of the three modes are $f_{1,\text{max}} = 4.217$ GHz, $f_{2,\text{max}} = 6.171$ GHz and $f_{3,\text{max}} = 7.578$ GHz.  To allow individual difference frequencies to be addressed, the mode frequencies are dispersed by modulating the impedance of the cavity along its length.}
\end{figure}

The device is a quarter-wavelength coplanar waveguide resonator \ref{Circuit} terminated by a SQUID at one end.  On the other end, it is capacitively overcoupled $\left(\text{Q}\approx7000\right)$ to a nominally $Z_0 = 50$ $\Omega$ line.  The device is made using Al and standard photolithography and e-beam lithography techniques. The fundamental mode has a relatively low frequency of around 1 GHz, giving higher modes with an average frequency spacing of 2 GHz, such that three higher-order modes are accessible within our 4-8 GHz measurement bandwidth.  Parametric processes are driven by a microwave pump inductively coupled to the SQUID, modulating the boundary condition of the resonator. Previous work demonstrated that this type of device could operate as a nondegenerate parametric amplifier operating near the standard quantum limit \cite{Simoen:2015by,Wustmann:2013bma,Wustmann:2017vka}. In a uniform cavity, the mode frequencies are equally spaced, making it difficult to address individual pairs of modes. To avoid this problem, we follow the approach of \cite{ZakkaBajjani:2011in} and modulate the impedance of the transmission line along the length of the cavity, varying the impedance from 41 to 72 $\Omega$. 

As has been well-documented \cite{Yamamoto:2008cr,Wilson:2010fj,Wilson:2011ir}, the SQUID parametrically couples the total flux in the cavity, $\hat{\Phi}_{c}$, to the pump flux, $\hat{\Phi}_{p}$ through its Hamiltonian  $\hat{H}_{\text{SQ}}=E_J|\cos(\pi\hat{\Phi}_{p}/\Phi_0)|\cos(2\pi\hat{\Phi}_{c}/\Phi_0)$ \cite{Johansson:2009hf}.
Starting from this relation, we can derive our interaction Hamiltonian by expanding to first order in $\hat{\Phi}_{p}$ (around the flux bias $\Phi_{\text{ext}}$) and to  second order in  $\hat{\Phi}_{c}$.  Further, applying the parametric approximation to the pump, we find
\begin{equation}
\ham_{\text{int}}=\hbar g_0\left(\alpha_{p}+\alpha_{p}^*\right)\left(\ann[a]{1}+\cre[a]{1}+\ann[a]{2}+\cre[a]{2}+\ann[a]{3}+\cre[a]{3}\right)^2	\label{Hint}
\end{equation}
where $\alpha_p$ denotes the coherent pump amplitude, the bosonic operators $\ann[a]{i},\cre[a]{i}$ correspond to the three cavity modes considered here, and $g_0$ is an effective coupling constant. Eq. (\ref{Hint}) contains a large number of terms corresponding to different physical processes. However, we can selectively activate different processes by the appropriate choice of pump frequency. For instance, by choosing the sum frequency $f_{p}=f_i + f_j$, $\hat{H}_{\text{int}}$ can be reduced to $H_{\text{DC}}=\hbar g\left(\ann[a]{i}\ann[a]{j}+\cre[a]{i}\cre[a]{j}\right)$ by using the appropriate rotating-wave approximation.  $H_{\text{DC}}$ is well-known to produce parametric downconversion, which creates (destroys) pairs of photons and has been used to produce entangled photons in a wide variety of systems.  In particular, it has been used in superconducting microwave systems to produce two-mode squeezing (TMS) \cite{Flurin:2012hq}, a form of CV entanglement.  If we instead choose to pump at the difference frequency $f_{p}=|f_i - f_j|$, $\hat{H}_{\text{int}}$ reduces to $\hat{H}_{\text{CC}}=\hbar g'\left(\ann[a]{i}\cre[a]{j}+\cre[a]{i}\ann[a]{j}\right)$. $\hat{H}_{\text{CC}}$ produces a coherent coupling between modes.  The internal cavity modes described by $\ann[a]{i}$ can be connected to the propagating modes exterior to the cavity, described by operators $\ann[a]{i,\text{o}}$, using standard input-output theory \cite{Wustmann:2013bma}.

In this Letter, we show experimentally that these two distinct classes of parametric processes can in fact be combined by simultaneously pumping at multiple frequencies, producing multipartite interactions between multiple modes in a way that is flexible and extensible. Because $H_{\text{DC}}$ and $H_{\text{CC}}$ do not commute with each other, it is not at all obvious that it should be possible to compose these operations in a straightforward manner.  In fact, the commutator of $H_{\text{DC}}$ and $H_{\text{CC}}$ plays an important role in generating the additional dynamics needed to generate multimode entanglement. This versatile method was first suggested in \cite{Bruschi:2016hu}, where it was shown that it is theoretically possible to produce multimode entangled states including CV cluster states.  Recent work has studied the computational complexity of the generated states, showing that they can be used for classically hard computations such as boson sampling \cite{Peropadre:2018dy}. The method generalizes previous work on squeezing \cite{Bruschi:2013cq}, mode-mixing quantum gates \cite{Bruschi:2013ka}, as well as entanglement \cite{Friis:2012ft,Friis:2013fn} in cavities undergoing relativistic motion. Earlier experimental work studied the development of multimode coherence in a parametric resonator pumped at two frequencies \cite{Paraoanu:2016fqa}.  To our knowledge, this is the first experimental work demonstrating that this scheme does, in fact, produce multimode entanglement, which has important implications for the field of relativistic quantum information among others.


We present two multipartite entanglement schemes, that we call the coupled-mode (CM) and bisqueezing (BS) schemes. Both  generate entanglement between three modes, but with a correlation structure that differs. In the CM scheme, the device is pumped simultaneously at $f_{p1}= f_1 + f_2$  and $f_{p2}=|f_3-f_1|$.  The pump at $f_{p1}$ produces TMS between $f_1$ and $f_3$,  while the pump at $f_{p2}$ coherently couples $f_3$ with $f_2$.  In the BS scheme \cite{Bruschi:2017jy}, the pump tones are applied at $f_{p1}=f_1 + f_2 $ and $f_{p2}=f_2+f_3$, directly producing TMS correlations between the pairs $f_1,f_2$ and $f_2,f_3$. 


We will characterize the entanglement in our propagating output states within the covariance formalism \cite{Simon:1994gb}.  With the good assumption that all of our $N$ modes are Gaussian \cite{Note1}, the state is fully characterized by the $2N \times 2N$ covariance matrix $\mathbf{V}$ of the I and Q voltage quadratures of the propagating modes. For the theoretical analysis, the measured voltage quadratures are calibrated and scaled, as shown below, to produce the quantities $\ann[x]{i}=\ann[a]{i,\text{o}}+\cre[a]{i,\text{o}}$ and $\ann[p]{i}=-i\left(\ann[a]{i,\text{o}}-\cre[a]{i,\text{o}}\right)$.  By collecting the N-mode quadrature operator terms into a vector operator $\mathbf{\hat{K}}=\left(\ann[x]{1},\ann[p]{1},\ann[x]{2},\ann[p]{2},\dots,\ann[x]{N},\ann[p]{N}\right)^T$, the elements in $\mathbf{V}$ are defined as $V_{ij}=\left\langle \hat{K}_i\hat{K}_j +  \hat{K}_j\hat{K}_i\right\rangle/2$ (assuming the modes are mean zero).

To test the validity of our calibration, we can first test if our measured covariance matrices are physical. To be physical in a classical sense, $\mathbf{V}$ has to be real, symmetric and positive semidefinite. 
To be physical in the quantum sense, $\mathbf{V}$ must also obey the Heisenberg uncertainty principle. 
It has been shown \cite{Simon:1994gb} that the uncertainty principle can be expressed in terms of the symplectic eigenvalues, $\nu_i$, of $\mathbf{V}$, which are found by diagnolizing $\mathbf{V}$ through a canonical transformation of $\mathbf{\hat{K}}$.  With these definitions, the uncertainty principle simply states $\nu_i \geq 1$ for all $i$. All the measured covariances matrices below were found to be physical according to these definitions \cite{SupNote}.

We can now study the entanglement properties of $\mathbf{V}$.  A common measure of entanglement in CV systems is the logarithmic negativity, $\mathcal{N}$, which derives from the positive partial transpose (PPT) criterion \cite{Simon:2000fd,Adesso:2005fw}. The physical picture of the PPT criterion is that if we time-reverse a subsystem (partition) of a multimode entangled state, then the resulting total state will be \textit{unphysical}.  Testing for entanglement then corresponds to confirming that the covariance matrix of the partial transpose state $\widetilde{\mathbf{V}}$ is \textit{unphysical}. That is, the entanglement condition is  $\widetilde{\nu}_{\text{min}}\equiv \nu_{\text{min}}(\widetilde{\mathbf{V}}) < 1$ or equivalently $\mathcal{N} \equiv \textrm{max}[0,-\ln(\widetilde{\nu}_{\text{min}})] > 0$.


The PPT criterion and $\mathcal{N}$ suffice to fully characterize two-mode Gaussian states but, as is well-known, classifying entanglement quickly grows complex with increasing $N$. Limiting ourselves to three-mode states, early work suggested classifying entanglement based on applying the PPT criterion to the three possible bipartitions of the state \cite{Giedke:2001ey, vanLoock:2003hn}.  This work proposed a highest class of ``fully inseparable" states, where all bipartitions are entangled. This class can be quantified by the so-called tripartite negativity $\mathcal{N}^{tri} = (\mathcal{N}^{A} \mathcal{N}^{B} \mathcal{N}^{C})^{1/3}$, where ${A,B,C}$ label bipartitions, which is only nonzero for fully inseparable states \cite{Sabin:2008ce}.

It was later pointed out \cite{Teh:2014ij,Shchukin:2015ci,Gerke:2016hu} that, although this test rules out that any one mode is separable from the whole, it does \textit{not} rule out that the state is a mixture of states, each of which is separable. That is, there exists states of the form $\rho = a \rho_1 \rho_{23} + b \rho_2 \rho_{13} + c \rho_3 \rho_{12}$, where $a + b + c = 1$, which are fully inseparable according to the above definition \cite{Teh:2014ij}.  It was suggested that the term ``genuine" tripartite entanglement be reserved for states that \textit{cannot} be written as such a convex sum.  We note that this distinction between full inseparability and genuine entanglement only exists for mixed states, so understanding the purity of the state under study is important.

Ref  \cite{Teh:2014ij} derived a set of generalized inequalities to test for genuine tripartite entanglement. We define linear combinations of our quadratures $u = h_1 x_1 + h_2 x_2 + h_3 x_3$ and $v = g_1 p_1 + g_2 p_2 + g_3 p_3$, where the $h_i$ and $g_i$ are arbitrary real constants to be optimized. It was shown that states without genuine entanglement satisfy the inequality 
\begin{equation}
S \equiv \langle \Delta u^2 \rangle + \langle \Delta v^2 \rangle \ge 2 \min \{| h_i g_i | + | h_j g_j + h_k g_k |  \}
\label{Bound}
\end{equation}
where the minimization is over permutations of $\{i,j,k\}$. We can reduce the optimization space and simplify the bound by putting restrictions on the coefficients. For this Letter, we will use the two cases i) $h_l = g_l = 1$, $h_m = h_n = h$, $g_m = g_n = g$, $h g < 1$ and ii) $h_l = g_l = 1$, $h_m = -g_n$, $h_n = -g_m$ both with the search domain $[-1,1]$. With these restrictions, the bound simplifies to 2.


To operate the device, the SQUID is flux biased to within 10\% of $\Phi_0$.  The pump tones are combined and feed to the on-chip pump line. The output of the device is fed through circulators to a cryogenic HEMT amplifier. After further amplification at room temperature, the signal is split in two paths and then fed through custom-made image-rejection filters into a pair of RF digitizers.  The digitizers output $I$ and $Q$ samples with a variable bandwidth.  In this work, the bandwidth was $BW = 1$ MHz. The variances and covariances of the $I$ and $Q$ time series are then computed. The measurements are done sequentially for the three mode pairs.  We remove the effects of drift by performing a chopped measurement, with the pumps turned on for 5 seconds followed by the pumps turned off for 5 seconds. The differenced data is then averaged over many cycles, typically 1000.

\begin{table*}
\begin{tabular}{|c||c|c||c|c|c|c|}
\hline
  &\multicolumn{2}{|c||}{Frequencies} & \multicolumn{3}{|c|}{Entanglement Measures} \\
\hline
Scheme & Modes &  Pumps & $\tilde{\nu}_{\text{min}}$ & $\mathcal{N}^{tri}$ & $S$\\
\hline
\hline
CM & 4.20, 6.16, 7.55 & 10.36, 3.35 & $0.48\pm0.002$, $0.39\pm0.002$, $0.57\pm0.002$ & $0.73\pm0.005$ & $1.49\pm0.01$\\
\hline
BS & 4.20, 6.16, 7.55 & 10.36, 11.75 & $0.31\pm0.003$, $0.48\pm0.004$, $0.39\pm0.004$ & $0.94\pm0.012$ & $1.19\pm0.01$\\
\hline
\end{tabular}
\caption{\label{Measures} Entanglement measures and frequencies for the various pumping schemes. 
CM is the coupled-mode scheme. BS is the bisqueezing scheme. The Frequencies columns list the respective mode and pump frequencies. The $\tilde{\nu}_{\text{min}}$ column reports the minimum symplectic eigenvalues for all three bipartition from the PPT tests. The $\mathcal{N}^{tri}$ column reports the tripartite negativity. The $S$ column reports the measure of genuine tripartite entanglement in Eq. 2.  The entanglement conditions are $\tilde{\nu}_{\text{min}}<1$;  $\mathcal{N} > 0$; and $S < 2$. Statistical errors are reported. See the Supplemental Material for a discussion of systematic error \cite{SupNote}. We find full inseparability and genuine tripartite entanglement for both entanglement schemes.}
\end{table*}

The entanglement tests described above compare the variances and covariances of the modes at the level of the vacuum noise.  It is therefore essential to have an accurate, absolute calibration of $\mathbf{V}$.  In this experiment, we perform this calibration using a shot noise tunnel junction (SNTJ) \cite{Spietz:2003vy,Spietz:2006ck} produced by NIST-Boulder \cite{SupNote}. 
 
The quadrature voltages at room temperature, $\Inp{i}$ and $\Qua{i}$, are converted to the scaled quadrature variables $\hat{x}_i$ and $\hat{p}_i$ using the calibrated system gains, $G_i$.  Following recent work \cite{Flurin:2015jf}, the scaled variance at the device output is
\begin{equation}
\langle\ann[x]{i}^2\rangle=\frac{4\left(\langle\Inp{i}^2\rangle_{ON}-\langle\Inp{i}^2\rangle_{OFF}\right)}{G_iZ_0 h f_i BW}+\coth\frac{h f_i}{2k_BT_i}
\end{equation}
with a similar definition for $\hat{p}_i$. The $\coth()$ term here represents the input quantum noise, at temperature $T_i$, which is (unfortunately) subtracted when we subtract the reference noise measured with the pump off. Without the input noise, the output variance will be underestimated, leading to an overestimate of the degree of entanglement or even an erroneous claim of entanglement.  It is therefore critical to characterize $T_i$ \cite{Bruschi:2013cd}.  Assuming the mode is in the vacuum state is tantamount to assuming that the system is entangled. In our setup, the calibration of the system gain using the SNTJ also gives us the physical electron temperature of the SNTJ.  As detailed in the Supplemental Material \cite{SupNote}, we find values of 25-37 mK over the course of our measurements.  For our working frequencies, these temperatures are deeply in the quantum regime, giving $\coth(h f_i/2k_BT_i) = 1.00$ with at least 3 significant figures for all our measurements.

Estimating the covariances of our modes is easier since neither the input noise nor system noise is correlated at different frequencies.  The covariance is then obtained by simply rescaling the room temperature values as, e.g.,
\begin{equation}
\langle\ann[x]{i}\ann[x]{j}\rangle=\frac{4\langle\Inp{i}\Inp{j}\rangle_{ON}}{\sqrt{G_iG_j f_i f_j}Z_0 h BW}.
\end{equation} 
As $\mathbf{V}$ is symmetric, just 21 terms in the matrix need to be individually measured for $N=3$ modes.

To study the tripartite CM scheme, we measure the 6 x 6 matrix $\mathbf{V}_{4,6,7}$:
\begin{align*}
	\bordermatrix{
		 & x_1 & p_1 & x_2 & p_2 & x_3 & p_3\ \\
    x_1 & 2.05&0.00&{\color{blue}1.87}&0.00&{\color{blue}0.88}&0.00\ \\
	p_1 & 0.00&2.04&0.00&{\color{red}-1.87}&0.00&{\color{blue}0.88}\ \\
	x_2 & {\color{blue}1.87}&0.00&2.85&0.00&{\color{blue}1.56}&0.00\ \\
	p_2 & 0.00&{\color{red}-1.87}&0.00&2.85&0.00&{\color{red}-1.56}\ \\
	x_3 & {\color{blue}0.88}&0.00&{\color{blue}1.56}&0.00&1.79&0.00\ \\
	p_3 & 0.00&{\color{blue}0.88}&0.00&{\color{red}-1.56}&0.00&1.79\ }.
\end{align*}
The correlations are color coded (online) with significant positive (negative) correlations in blue (red). As reported in Table \ref{Measures}, this state demonstrates both full inseparability and genuine tripartite entanglement.  This is the major result of this Letter.

For the BS scheme, we have the measured matrix $\mathbf{V}_{4,6,7}$:
\begin{align*}
	\bordermatrix{
		& x_1  & p_1 & x_2  & p_2  & x_3  & p_3\ \\
    x_1 & 3.91&0.00&{\color{blue}2.34}&0.00&{\color{blue}2.78}&0.00\ \\
	p_1 & 0.00&3.91&0.00&{\color{red}-2.33}&0.00&{\color{red}-2.78}\ \\
	x_2 & {\color{blue}2.34}&0.00&2.28&0.00&{\color{blue}1.45}&0.00\ \\
	p_2 & 0.00&{\color{red}-2.33}&0.00&2.28&0.00&{\color{blue}1.45}\ \\
	x_3 & {\color{blue}2.78}&0.00&{\color{blue}1.45}&0.00&2.72&0.00\ \\
	p_3 & 0.00&{\color{red}-2.78}&0.00&{\color{blue}1.45}&0.00&2.72\ }.
\end{align*}
As shown in Table \ref{Measures}, we again find that the state demonstrates both full inseparability and genuine tripartite entanglement. 

The main limitation on the degree of entanglement in the system seems to be the purity of the output states.  For an ideal system, pumping harder should increase the degree of squeezing without degrading the purity of the state, and therefore increase the parametric gain and degree of entanglement monotonically.  We instead see that the gain increases with pump strength, but the purity of the states simultaneous declines, limiting the maximum degree of entanglement. This suggests some form of nonideality such as higher-order nonlinearities, self-heating, or parasitic coupling to other cavity modes. These limitations can be more thoroughly investigated in future work.

The authors wish to thank B. Plourde, J.J. Nelson and M. Hutchings at Syracuse University for invaluable help in junction fabrication.  We would also like to acknowledge M. Piani for helpful discussions. CMW, CWSC, PFD and AMV acknowledge NSERC of Canada, the Canadian Foundation for Innovation, the Ontario Ministry of Research and Innovation, Canada First Research Excellence Fund (CFREF), Industry Canada, and the CMC for financial support. Financial support by Fundaci{\'o}n General CSIC (Programa ComFuturo) is acknowledged by CS as well as additional support from Spanish MINECO/FEDER FIS2015-70856-P and CAM PRICYT Project QUITEMAD+ S2013/ICE-2801. IF acknowledges support from EPSRC (CAF Grant No. EP/G00496X/2). FQ and GJ acknowledge the support from the Knut and Alice Wallenberg Foundation.

\bibstyle{apsrev}
\bibliography{CoupledModePaper,Notes}

\end{document}